\documentclass{aa}
\usepackage{graphicx}
\usepackage{txfonts}
\usepackage{lscape}

\begin{document}
   \title{Infrared identification of 2XMM J191043.4$+$091629.4}

   \author{J. J. Rodes-Roca\inst{1,2}, J. M. Torrej\'on\inst{1,2},
          S. Mart\'{\i}nez-N\'u\~nez\inst{2}, G. Bernab\'eu\inst{1,2},
          \and
          A. Magazz\'u\inst{3}
          }

   \offprints{J. J. Rodes-Roca}

   \institute{Department of Physics, Systems Engineering and Signal Theory, University of Alicante,
              03080 Alicante, Spain\\
              \email{rodes@dfists.ua.es}\\
         \and
              University Institute of Physics Applied to Sciences and Technologies,
              University of Alicante, 03080 Alicante, Spain\\
         \and
Telescopio Nazionale Galileo, Rambla Jos\'e Ana Fern\'andez P\'erez, 38712 Bre\~na Baja, Spain\\
                          }

   \date{Received     ; accepted       }

  \abstract
   {We report the infrared identification of the
   X-ray source 2XMM J191043.4$+$091629.4, which was detected by
   \emph{XMM-Newton/EPIC} in the vicinity of the
   Galactic supernova remnant W49B.}
   {The aim of this work is to establish the nature of the X-ray source
   2XMM J191043.4+091629.4 studying both the infrared photometry and
   spectroscopy of the
   companion.}
   {We analysed UKIDSS images around the best position of the X-ray source
   and obtained spectra of the best candidate using \emph{NICS} in the
   \emph{Telescopio Nazionale Galileo} (TNG) 3.5-m telescope.
   We present photometric and spectroscopic TNG analyses of the infrared
   counterpart of the X-ray source,
   identifying emission lines in the $K$-band. The $H$-band spectra does not present any
   significant feature.}
   {We have shown that the Brackett $\gamma$ H \textsc{I} at 2.165 $\mu$m, and
   He \textsc{I} at 2.184 $\mu$m and at 2.058 $\mu$m are significantly present
   in the infrared spectrum. The CO bands are also absent from our spectrum.
   Based on these results and the 
   X-ray characteristics of the source, we conclude that the infrared counterpart
   is an early B-type supergiant star with an $E(B-V)= 7.6\pm 0.3$
   at a distance of 16.0$\pm$0.5 kpc. This would be, therefore, the
   first high-mass X-ray binary in the Outer Arm at galactic longitudes
   of between 30$^\circ$ and 60$^\circ$.}
   {}

   
   \authorrunning{J. J. Rodes-Roca et al.}
   \titlerunning{Infrared identification}
   \maketitle
%

\section{Introduction}

2XMM J191043.4+091629.4 is an unclassified X-ray source
detected serendipitously by \emph{XMM-Newton} in the
vicinity of the Galactic supernova remnant W49B (2XMM; \cite{2XMM} 2009).
This X-ray source
was discovered with \emph{ASCA}, AX J1910.7+0917, during the survey of the
Galactic plane (\cite{sugizaki}, who associated it to the \emph{Einstein}
source 2E 1908.3+0911). The source was detected by the
\emph{INTErnational Gamma-Ray Astrophysics Laboratory} (INTEGRAL;
\cite{winkler03}) in the imager \emph{ISGRI} during an observation
of the SGR 1900+14 field (\cite{gotz06}, see Fig. 1 in this
reference).

\emph{INTEGRAL}, \emph{ASCA}, \emph{XMM-Newton}, and \emph{Chandra}
observations suggested that the source could be a binary transient system
associated with the IR counterpart 2MASS J19104360+0916291.
The X-ray spectrum seems to be characteristic of a high-mass
X-ray binary (HMXB) system with a neutron star as a compact object
(although no pulsations have been detected so far;
see \cite{pavan11} 2011).
However, neither a classical
supergiant wind-fed system nor a Be/X-ray binary fit
the observed behaviour well according to \cite{pavan11} (2011).
These authors used the \emph{XMM-Newton} position of the
X-ray source to pinpoint the IR counterpart
using the 2MASS catalogue and argued that the photometric colours
favour a supergiant companion. However, in a deeper search,
we found two possible near-IR
counterparts in the UKIDSS-GPS DR5 catalogue
(United Kingdom Infrared Deep Sky Survey--Galactic
Plane Survey: Data Release 5) that were astrometrically
coincident with 2.13$^{\prime\prime}$ \emph{XMM-Newton} error circle. These
two sources appeared unresolved in the 2MASS images that were
identified with the 2MASS J19104360+0916291 source
(see Fig. 7 in \cite{pavan11} 2011 and Fig. 1 in \cite{Rodes11}).
This means that the 2MASS photometry is contaminated. 
We have performed a photometric study of the possible counterparts
and have found that the candidate \#1 in Fig.~\ref{counterpart}
is the most likely one. We also obtained reliable
photometry.

The source is located in the Galactic plane and has a relatively
high absorption in the X-ray domain ($N_H \sim 5\times 10^{22}$
cm$^{-2}$, \cite{pavan11} 2011). The lack of an obvious optical
counterpart (\cite{pavan11} 2011) is also compatible with these
characteristics. To advance our current understanding on the
nature of AX J1910.7+0917, we used observations in the near-IR
range acquired with the \emph{Telescopio Nazionale Galileo}
(TNG) 3.5-m telescope.
The characterization of the IR counterpart of 2XMM J191043.4+091629.4 is particularly challenging. On one hand the spectral classification of hot stars based on a $K$-band spectrum cannot
be completed without ambiguities because of the lack of
enough spectral features in that range (\cite{hanson96} 1996).
This problem can be circumvented by combining data for
several spectral bands, including the X-rays. On the other hand this system lies in the line of sight of a second (unrelated) star, making it a visual binary with a separation of only 1$^{\prime\prime}$. This, together with the strong absorption, requires the use of a 4-m class telescope under very good seeing conditions.

In the framework of an ongoing programme to discover and
characterize optical counterparts to HMXBs, we have studied
this source. Here, we present observations of the TNG
using the Near-Infrared Camera Spectrometer (NICS). According to the
near-IR spectral
and photometric
properties, we propose that the counterpart is most likely an early-type B supergiant
star.

\section{Observations and data reduction}
\label{data}

\subsection{The \emph{XMM-Newton} position and the IR candidate}

\begin{figure*}[htb]
  \centering
  \includegraphics[angle=0,width=\columnwidth]{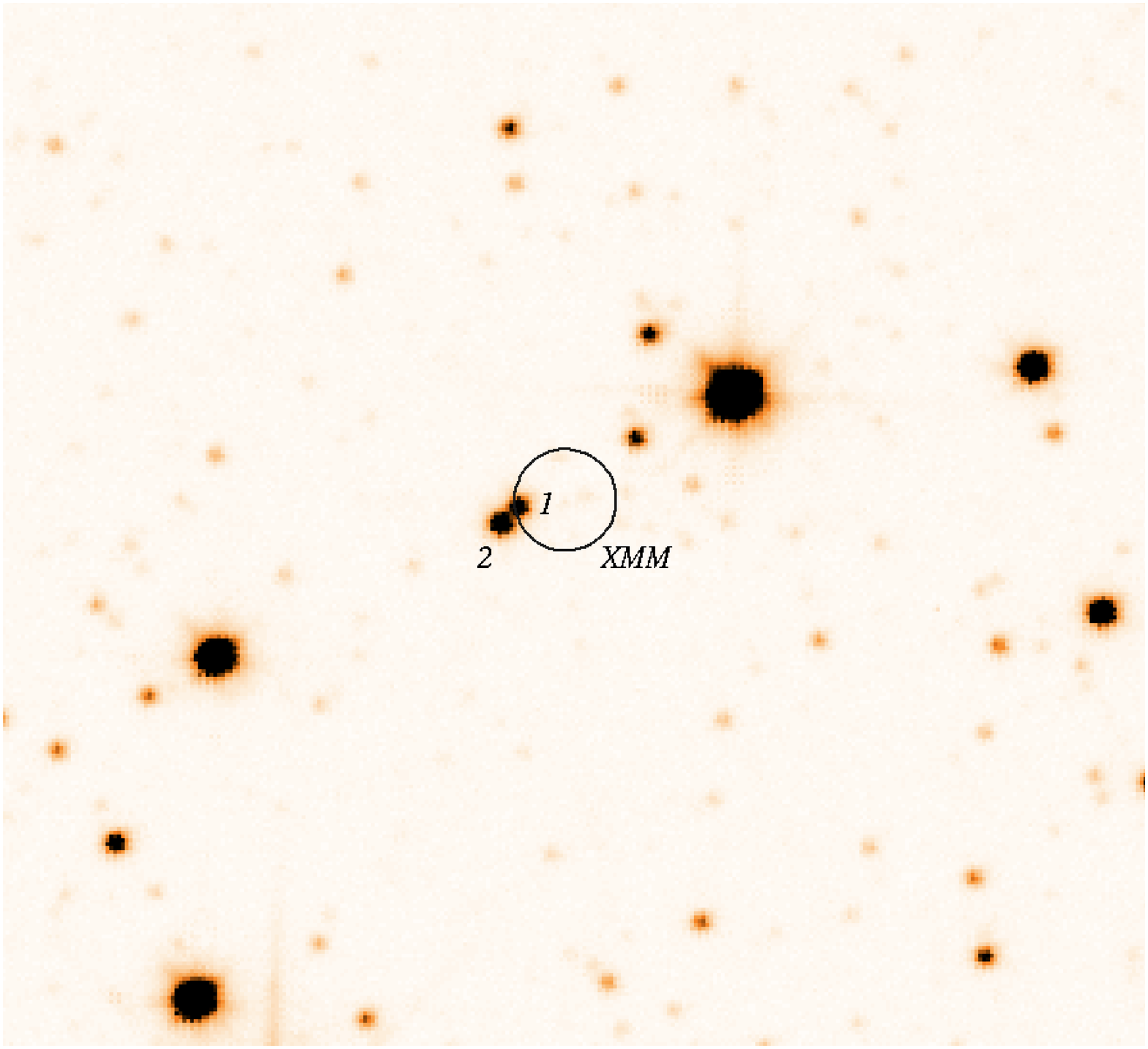}
  \includegraphics[angle=0,width=\columnwidth]{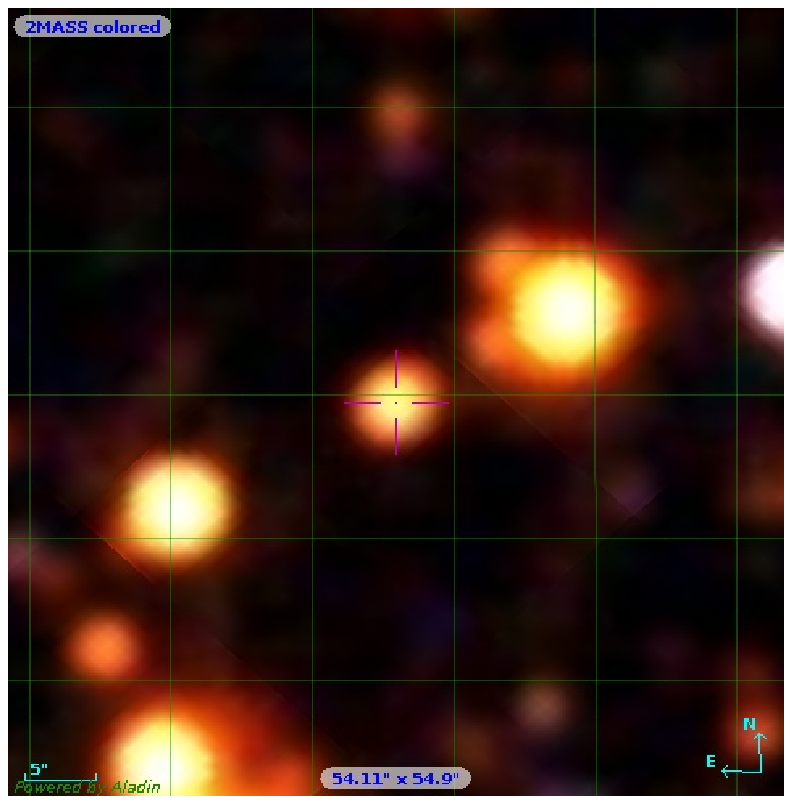}
  \caption{\emph{Left panel}: 15$^{\prime\prime}$ $\times$15$^{\prime\prime}$ \emph{K} finding chart for 2XMM J191043.47$+$091629.4.
  The black circle is centred on the \emph{XMM-Newton} position of 2XMM
  J191043.47$+$091629.4, with the radius indicating the 2.13$^{\prime\prime}$
  positional error.
\emph{Right panel}: 3.6'$\times$2.6' 2MASS coloured map. The images
are displayed with north up and east to the left. We note that the
two NIR UKIDSS sources appear unresolved in the 2MASS image.
  }
  \label{counterpart}
\end{figure*}

UKIDSS is a NIR survey covering
approximately 7000 deg$^2$ of the northern hemisphere to a depth of
$K = 18$ mag, with additional data from two deeper, small-area high-redshift
galaxy surveys. Using the Wide Field Camera (WFCAM) on the United Kingdom
Infrared Telescope (UKIRT), the survey achieved a pixel resolution of
0.14$^{\prime\prime}$
by use of the micro-stepping technique (see \cite{lawrence07}, for full details).
The data used in this paper were taken from the UKIDSS-GPS, a survey of
approximately 2000 deg$^2$ of the northern Galactic plane in the \emph{J},
\emph{H} and \emph{K}-bands (\cite{lucas08}).
In Figure~\ref{counterpart} we show the UKIDSS DR7PLUS Galactic plane survey
\emph{K}-band image
of the sky around the position of the X-ray source. As can be seen,
the superior spatial resolution of the UKIDSS survey images is able
to resolve the 2MASS candidate into two different components,
labelled here as \#1 and \#2. We have also plotted the error circle
centred at the best position, namely, $\alpha$ = 19$^h$ 10$^m$ 43.40$^s$
and $\delta =$ +09$^\circ$ 16$^\prime$ 30.0$^{\prime\prime}$,
with an error of 2.13 arcsec
(\cite{pavan11} 2011). As can be seen, candidate \#2 lies completely
outside the error circle and only candidate \#1 is compatible with
the \emph{XMM-Newton} position.

\subsection{TNG data}

Near-IR spectroscopy was obtained
from 14-16 July 2012, using the NICS spectrometer at the 3.5-m TNG
telescope.
The scientific and calibration
data were retrieved from the Italian Centre for
Astronomical Archive (IA2). Low-dispersion spectroscopy was obtained
with the HK grism, which covers the 1.40--2.50 $\mu$m
spectral region and provides a dispersion of 11.2 $\AA$/pixel.
We note that the component separation is around 1 arcsec, making both
spectroscopy and photometry very challenging. For that reason we selected only nights with excellent seeing conditions of 0.75$^{\prime\prime}$ or less. Furthermore, we used a slit of 0.5$^{\prime\prime}$ and oriented it 
perpendicularly to the line of union of the two stars
to acquire only
light from the correct candidate, labelled as \#1 in
Fig~\ref{counterpart}.

To remove the sky background, the source and the standard stars
were observed at several positions along the slit following a dithering ABBA sequence. Consequently, the
observation consisted of a series of four images with the source spectrum displaced at different positions along the CCD, using an automatic script available at the telescope.
First, cross-talking effects produced by
non-saturated images were corrected using the Fortran program
available for this purpose.
Second, background subtraction was made
obtaining our (A$-$B) and (B$-$A) image differences.
The sequences AB and BA were so close in time
that sky background variation between them was negligible.
This method, together with the use of the 0.5$^{\prime\prime}$ slit, also
minimizes any possible nebular contamination. Third,
sky-subtracted images were flat-fielded and the resulting
spectra averaged. The wavelength calibration was performed using
the Ar lamps available at the telescope. Fitting a third-degree
polynomial, the root mean squared error was around 1.7 $\AA$.
Fourth, as telluric calibration source we used A0 V Hip98640
star because
this type of stars is featureless in the K band.
Then, we normalized
both spectra considering the K band from 2.05 $\mu$m to 2.2 $\mu$m
and the H band form 1.6 $\mu$m to 1.75 $\mu$m by dividing
the source spectrum by the standard spectrum
to identify the spectral lines of the source.
Finally, to minimize the noise, we filtered the high
frequencies above the Nyquist frequency
$\sigma_N = 1/(2\cdot\Delta x) = 0.0025$ $\AA^{-1}$ in the Fourier
transform of the spectrum, recovering the cleaned spectrum
by computing the inverse Fourier transform.
The final K-band spectrum is shown in Fig.\ref{Kband}.

\section{Data analysis}
\label{analyse}

\subsection{Near-IR spectrum and classification of the counterpart}

The spectral analysis was carried out using the
\emph{Starlink}\footnote{http://starlink.jach.hawaii.edu/starlink} software.
To identify the emission/absorption lines and
spectral classification, we used the following atlases:
\cite{blum97}, \cite{meyer98} and \cite{hanson98} for
the H band; \cite{hanson96} (1996) and \cite{hanson05} (2005) for
the K band.

\begin{figure}[htb]
  \centering
  \includegraphics[angle=-90,width=\columnwidth]{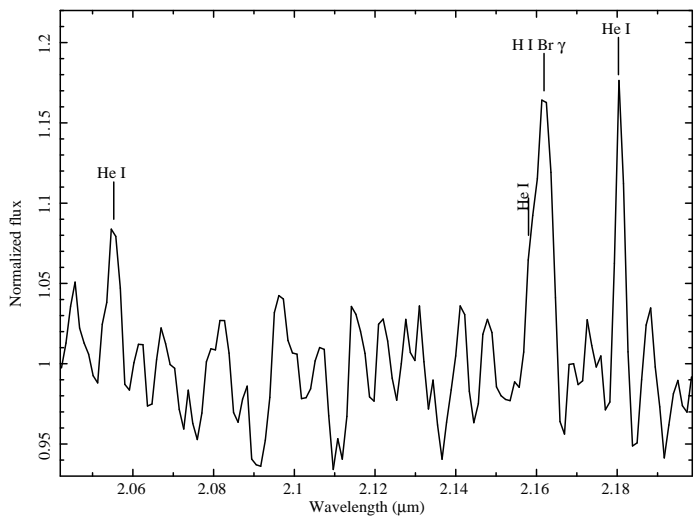}
  \caption{K-band spectrum of the counterpart to the X-ray
binary studied in this work. Note  the absence of any feature
at the position of the He \textsc{ii} 2.1885 $\mu$m line.
  }
  \label{Kband}
\end{figure}

Figure~\ref{Kband} shows the K-band spectrum
of the IR counterpart. The presence of He \textsc{i} lines
and the absence of He \textsc{ii}, which is seen up to O9,
points towards a B-type star. The Br$\gamma$ line is
in emission and seems to be blended with the blueward
emission of the He \textsc{i} 2.161 $\mu$m line. We also
observe He \textsc{i} 2.058 $\mu$m in emission as in
B supergiants and Be stars. In the atlas of \cite{hanson05}
(2005), no luminosity class III star shows this line
in emission.  Moreover, we also
note the presence of the He \textsc{i} 2.183 $\mu$m
emission line, which becomes apparent in early-B giants, albeit
in absorption (see Figs. 5 and 8 to 12 in \cite{hanson05}
2005).

Unfortunately, the low signal-to-noise ratio of the data
in the H-band spectrum
prevented us from identifying hydrogen lines
such as Br10, Br11, or Br12, and/or helium lines
like He \textsc{ii} 1.6918 $\mu$m or He \textsc{i} 1.7002 $\mu$m
clearly. Other hydrogen lines
are not detected, probably because their intensities would be
below the continuum noise level.
On the other hand, cool supergiant stars show CO-band
absorption lines between 2.29 and 2.35 $\mu$m
that are not present in our spectrum.
We therefore rule out a late-type companion for
2XMM J191043.4+091629.4.

In conclusion, the spectral type corresponds to an early-B star while
the luminosity class can not be constrained from the near-IR
spectrum and is consistent with either class I (supergiant) or class V (Be star).
In the following, however, we argue that the
X-ray characteristics of the source are more compatible with
a supergiant X-ray binary system.

\subsection{IR photometry}
\label{IR}

\begin{figure}[htb]
  \centering
  \includegraphics[angle=0,width=\columnwidth]{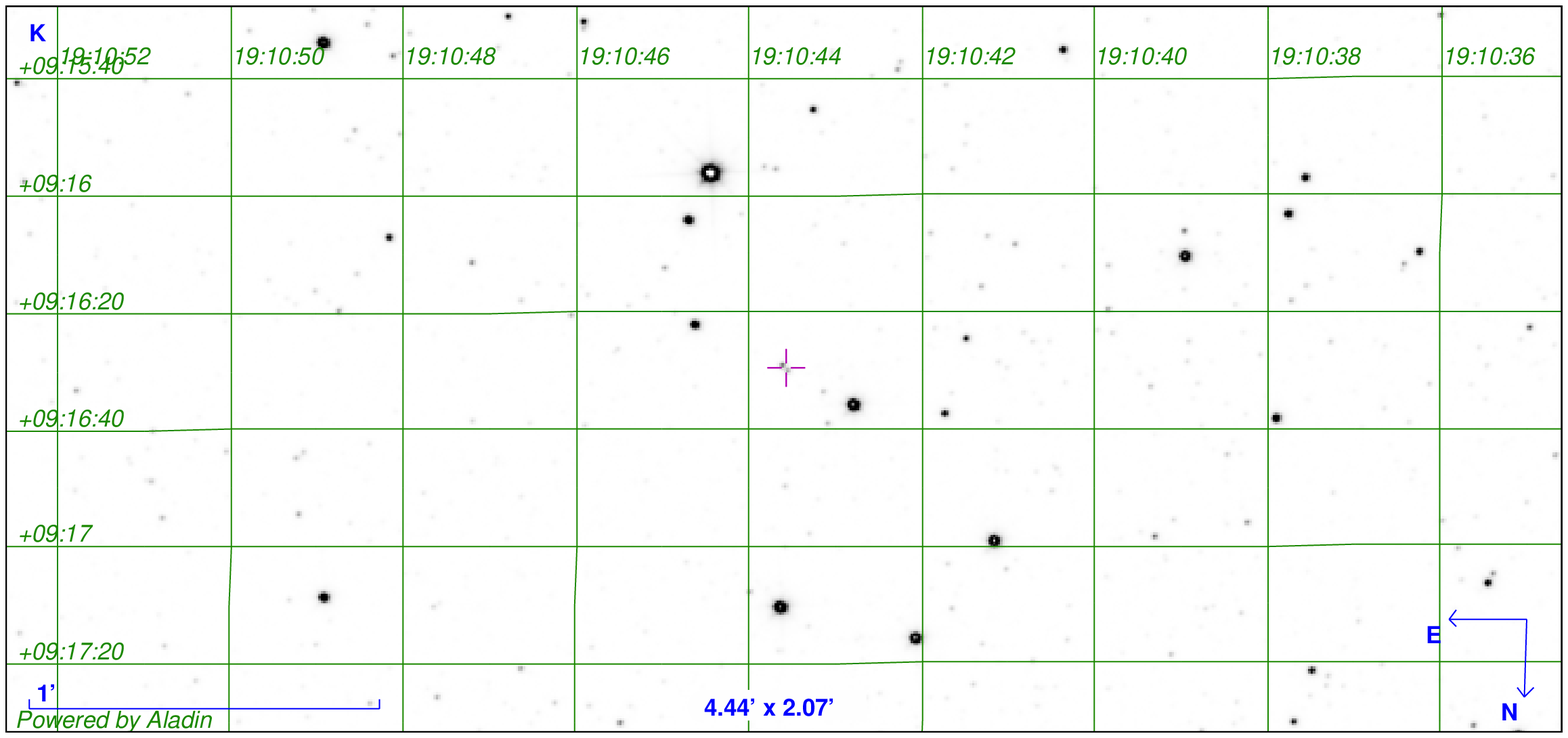}
  \includegraphics[angle=0,width=\columnwidth]{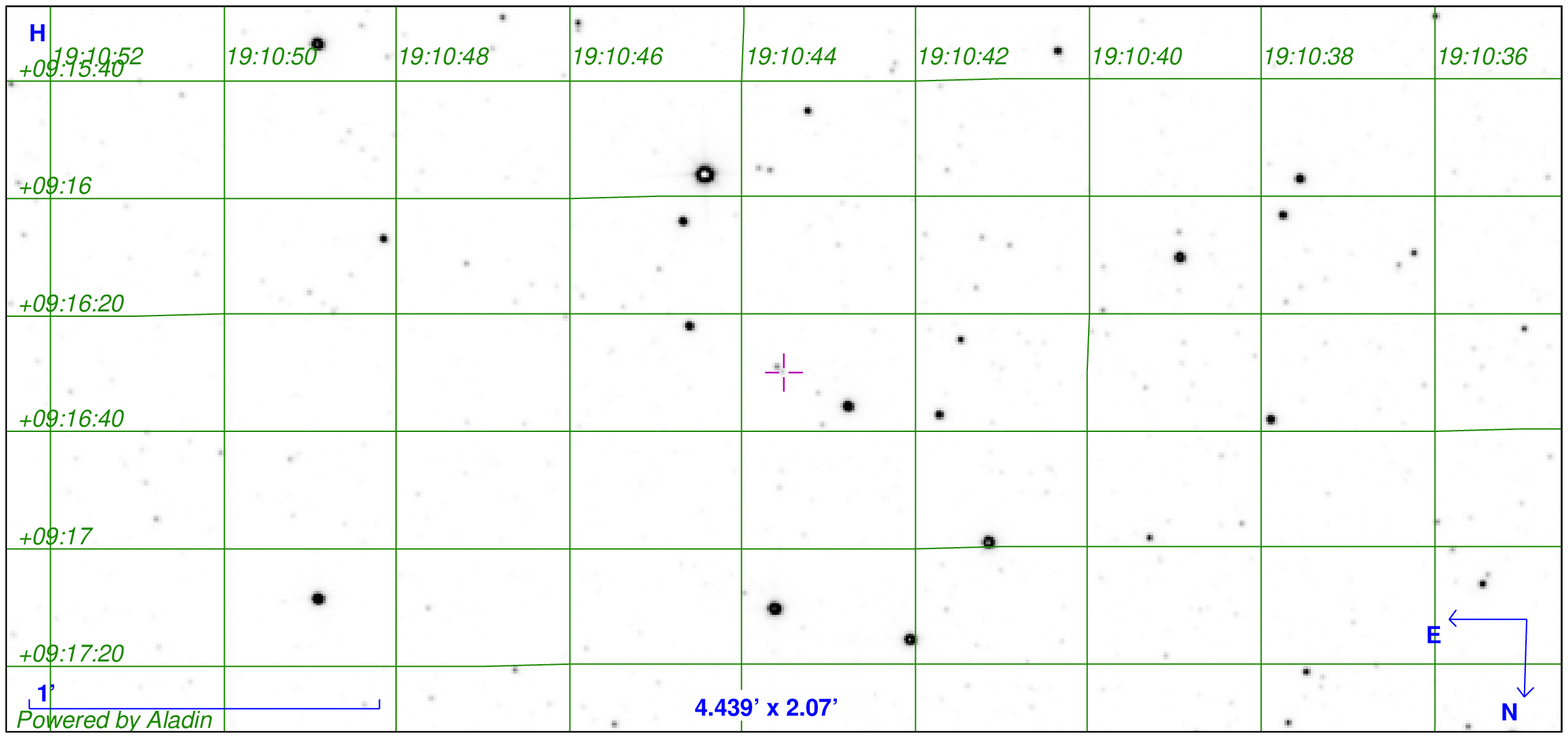}
  \includegraphics[angle=0,width=\columnwidth]{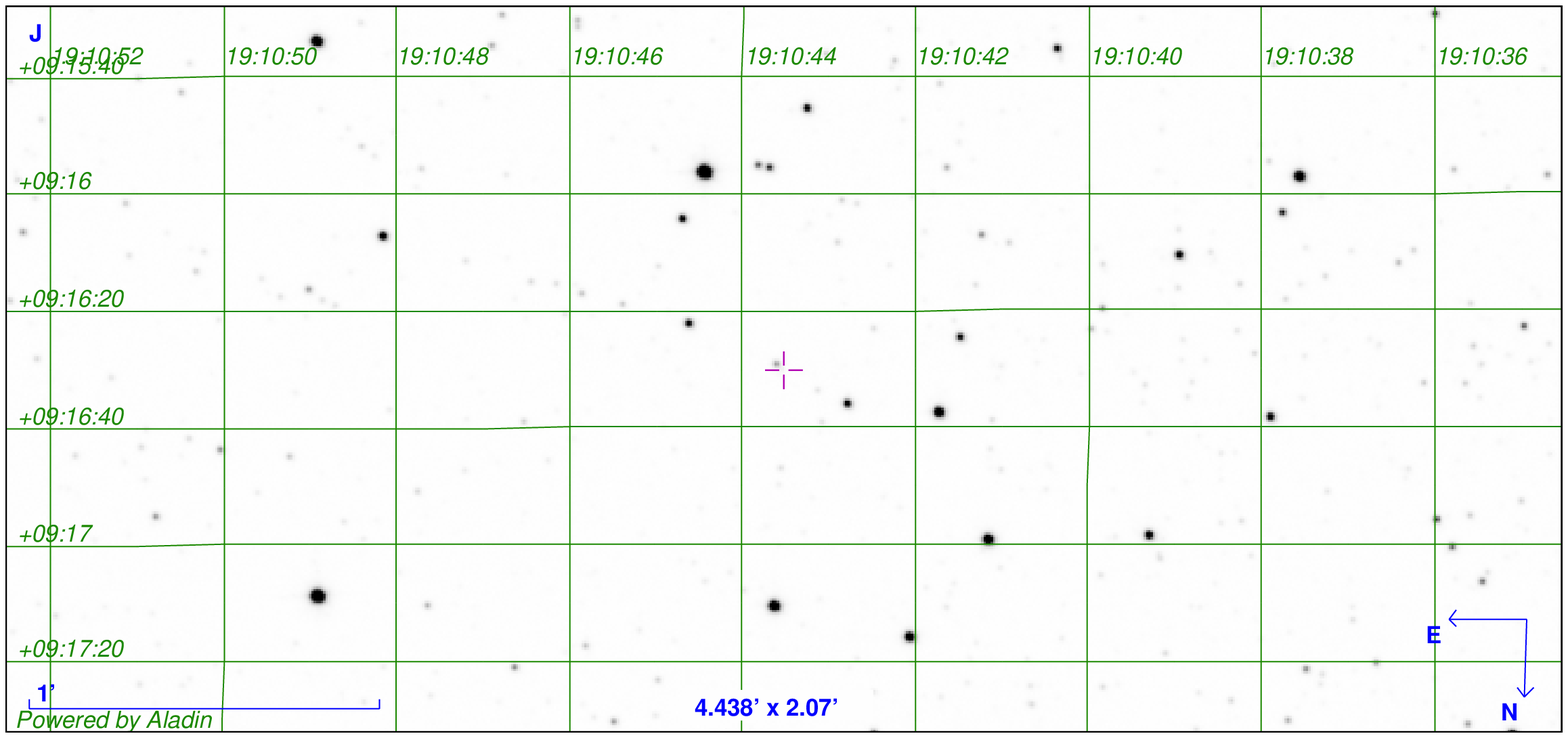}
  \caption{4.4'$\times$2.1'
  \emph{K} (top), \emph{H} (middle),
  and \emph{J} (bottom) finding chart for 2XMM J191043.47$+$091629.4.
  The cross is centred on the \emph{XMM-Newton} position of 2XMM
  J191043.47$+$091629.4. The UKIDSS image is displayed with north
down and east to the left.
  }
  \label{ukidss}
\end{figure}

To carry out the photometric analysis of the images we used
the \emph{Starlink} software, in particular the Graphical
Astronomy and Image Analysis Tool (\emph{GAIA}) package.
As shown in previous sections, the 2MASS candidate is actually an
unresolved pair. Therefore the 2MASS photometry for this candidate
is contaminated and cannot be used to compute the distance to the
source. We used the UKIDSS images instead (see Fig.~\ref{ukidss}).
In the UKIDSS database
only the $K$ magnitude for candidate \#1 is available ($K=13.135\pm 0.003$),
because the counterpart is very weak and the automatic extraction
only gives a poor solution. However, the counterpart is clearly visible
also in $H$- and $J$ images, although barely in this last band.
To perform the photometry, we extracted the fluxes of candidate \#1
as well as those of several dozens of other stars seen in the image using synthetic
aperture photometry. Background fluxes were also extracted from
source-free regions in the same image, close to the different stars. The
instrumental magnitudes were then correlated with the corresponding
magnitudes available at the UKIDSS database. Finally, the calibration
equations were applied to the fluxes of candidate \#1 to obtain the photometry
given in Table~\ref{photometry}.

\begin{table}
\begin{tabular}{ccc}
\hline
\hline
$J$ & $H$ & $K$ \\
\hline
$\gtrsim 17.1$ & 14.43$\pm 0.05$ & 13.135$\pm$ 0.003 \\
\hline
\hline
\end{tabular}
\caption{Photometry of candidate \#1 in the UKIDSS system.}
\label{photometry}
\end{table}

As a consistency check, the $K$ magnitude obtained from the direct
application of the previous calibrated equations to source \#1 was identical,
within the errors, to that quoted in the UKIDSS database. Therefore,
the magnitudes and errors given in Table~\ref{photometry} are reliable. No error
is given for the $J$ magnitude because source \#1 is already very
weak in this band and only an upper limit can be obtained. Clearly,
the source is strongly reddened.

Assuming a B0I type, the intrinsic colour would be $(H-K)_{0}=-0.08$
(\cite{ducati01}). Now using the photometry in
Table~\ref{photometry}, we can estimate an infrared excess
of $E(H-K)=1.30$. This corresponds to an $E(B-V)=E(H-K)/0.17=7.6\pm 0.3$
(\cite{fitz99}). This high value agrees
with  the column density deduced from the X-ray analysis
($N_{\rm H}=6\times 10^{22}$ cm$^{-2}$; \cite{pavan11} 2011),
which would correspond to an $E(B-V)=8.8$
(\cite{ryter96}). This last value is
obtained assuming that the entire column is interstellar but,
in fact, part of it will be local if the compact object is
embedded in the companion's wind. Note that a later
spectral type would reduce the value of the
total reddening $E(H-K)$ still more, and owing to the spectral
analysis of the previous section, we can rule out a
late-type star. The available data then
are more consistent with an early-type companion.
The total-to-selective
absorption will be $A_{K}=0.36 \; E(B-V)=2.74$. Now, assuming an
absolute magnitude $M_{K}=-5.6$ for a B0I star, this would
translate into a distance to the source of $d=16.0\pm 0.5$ kpc. 

On the other hand, Br$\gamma$ is the most prominent feature in Be stars in the
$K$ band, while He \textsc{i} 2.058 $\mu$m is found in early-type
Be stars, up to B2.5 (\cite{CS2000}). Therefore, assuming a B0V star, the intrinsic colour
would be $(H-K)_{0}=-0.05$ (\cite{ducati01}) and $E(H-K)=1.27$,
also compatible with the X-ray column density. The $M_{K}=-3.17$
and the corresponding distance $d=5.3$ kpc. From the analysis
of X-ray data (\cite{pavan11} 2011), the average unabsorbed flux is
of the order of $2\times 10^{-11}$ erg s$^{-1}$ cm$^{-2}$. At a
distance of 5.3 kpc this would translate into an X-ray luminosity
of $L_{X}=6.7\times 10^{34}$ erg s$^{-1}$. This is two to three orders
of magnitude lower than the typical luminosities for type I
outbursts in transient BeX-ray binaries (BeXBs).
On the other hand, there is
a growing class of systems called persistent BeXBs that
are characterized by low X-ray luminosities,
$L_{(2-20) \; \; keV} \sim 10^{34-35}$ erg s$^{-1}$ (\cite{pau2011}).
These systems are relatively
quiet, showing flat light curves with sporadic
and unpredictable increases in intensity by less than one
order of magnitude, and very weak, if any, iron
fluorescence line at $\sim$6.4 keV. The upper limit on
the X-ray flux in quiescence of the source AX J1910.7+0917 (1--10 keV energy band)
is $5.4\times 10^{-11}$ erg s$^{-1}$ cm$^{-2}$ (
\cite{pavan11} 2011 from a \emph{Chandra}
observation), implying an X-ray luminosity during quiescence of
$L_{(1-10) \; \; keV}\lesssim 1.8\times 10^{33}$ erg s$^{-1}$, which is below the limit
displayed by the persistent BeXBs discovered so far.
Therefore, the X-ray observations
are inconsistent with
a classical transient BeXB, but could be compatible with a
low-luminosity persistent BeXB.

Finally, the K-band spectrum, which
shows narrow emission lines due to He \textsc{i} and Br$\gamma$
(\cite{howell10}), would be compatible with those displayed by CVs.
But, again, this possibility is ruled out by the X-ray emission
characteristics from
AX J1910.7+0917 as discussed in \cite{pavan11} (2011).

\begin{figure}[htb]
  \centering
  \includegraphics[angle=0,width=\columnwidth]{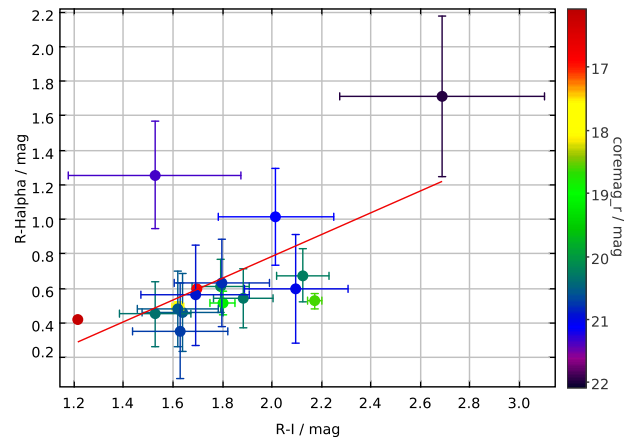}
  \caption{Colour-colour diagram of the IPHAS detections (not
sources) in the 1.3 arcmin field around \emph{XMM-Newton} coordinates.
Objects with H$\alpha$ excess are located towards the top of
the diagram. The $r$ magnitude is colour-coded.
  }
  \label{halpha}
\end{figure}

We also searched for a possible H$\alpha$ emission from the system
using Isaac Newton Telescope
Photometric H$\alpha$ Survey (IPHAS, \cite{drew05}).
We followed the Euro-Virtual Observatory (VO)
scientific case developed by \cite{ZCh11}. We were able to select 23
IPHAS sources inside a 1.3 arcmin circular field around
the \emph{XMM-Netwon} best position. Using VO tools, namely
\textsc{CDS ALADIN} (\cite{aladin}) and \textsc{TOPCAT}
(\cite{topcat}), we explored the colour-colour diagram
of the IPHAS sources (see Fig.~\ref{halpha}).

The straight line in Fig.~\ref{halpha} roughly corresponds to
main-sequence stars, which do not exhibit H$\alpha$ emission, while
outlier points correspond to H$\alpha$ emitters. We found
a single detectable prominent H$\alpha$ emitter whose coordinates
$\alpha=19^h 10^m 42.94^s$ and
$\delta=+09^\circ 16^{\prime} 01.6^{\prime\prime}$
are outside the 2.13$^{\prime\prime}$ \emph{XMM-Newton}
error circle, however
(the difference between the coordinates is
$\Delta\alpha=10.4^{\prime\prime}$ and $\Delta\delta=28.4^{\prime\prime}$,
implying an angular separation from the XMM source position of 30.2$^{\prime\prime}$).
The non-detection of H$\alpha$ emission from the system is not strange, though.
In the previous section, we derived a reddening value of $E(B-V)=7.6$,
implying an extremely high extinction in $R$ ($A_R\sim16$ mag)
and $I$ ($A_I\sim12-13$ mag). It is therefore expected that IPHAS
has no detections inside the \emph{XMM-Newton} error circle
since the object is already very faint in $J$.

In conclusion, the most likely counterpart to 2XMM J191043.4+091629.4 is
an early-type B I star located at a distance\footnote{The error would be $\pm 0.3$ kpc if
we assumed a 0.5 magnitude error in the absolute magnitude calibration
for early-type supergiants} of
$d=16.0\pm 0.5$ kpc, placing this source in the Outer Arm.

\section{Summary and discussion}
\label{conclusion}

According to the previous analysis, we have identified
the counterpart to 2XMM J191043.4+091629.4 as being a
B supergiant star. This implies that 2XMM J191043.4+091629.4
belongs to the class of obscured HMXBs that has been
unveiled in the past decade by satellites such as \emph{INTEGRAL}. A large
fraction show intrinsically high absorption
along the line of sight ($N_H \gtrsim 10^{23}$ cm$^{-2}$
\cite{kuulkers}). These IGR sources share similar X-ray
properties typical for
accreting X-ray pulsars in HMXBs. The unabsorbed X-ray
luminosity in the energy range 2--100 keV in of the order
10$^{35}$--5$\times$10$^{36}$ erg s$^{-1}$ is also typical for HMXBs.
In addition, the vast majority of these newly discovered
sources had supergiant donors.

Assuming a distance to the source 2XMM J191043.4+091629.4
of 16.0 kpc (see Sect.~\ref{IR}) and isotropic emission,
the estimated X-ray luminosity is (0.9--1.3)$\times$10$^{36}$
erg s$^{-1}$ in the 0.3--10.0 keV range. Moreover, the photon
index $\Gamma \sim 1.2$, and the $N_H \sim 0.5\times10^{23}$
cm$^{-2}$ are also consistent with the characteristics of
the new population of highly absorbed supergiant HMXBs. This is compatible also with the lower band of the typical X-ray
luminosity of {\it classical} supergiant X-ray binaries (SGXBs) such as Vela X-1
or 4U 1538$-$52, which is
of the order of $L_{X}\simeq 10^{36}$ erg s$^{-1}$.

Moreover, the source is heavily obscured, with an $E(B-V)=7.6$
implying extinctions of about $A_V\sim23.6$ mag
in the visual band. At an estimated distance of 16.0 kpc, the source would be located in the
Outer Arm.  The line of sight, then, crosses the heavily populated Perseus arm and, perhaps, the Sagittarius arm tangent, which explains the high extinction displayed by the system.
On the other hand, with the data at hand
we cannot discard completely
a persistent BeXB which, located at
$d=5.3$ kpc, would be the faintest ($L_{X}\simeq 10^{33}$ erg s$^{-1}$)
found so far, however.

Until \emph{INTEGRAL} discovered supergiant fast
X-ray transients (SFXTs) and highly absorbed supergiant X-ray
binaries (SGXBs), the population of these systems
was relatively small, in agreement with evolutionary scenarios.
Most of them were known because they were
persistent, moderately bright X-ray sources.
These class of systems are growing and,
currently, \emph{INTEGRAL} has discovered
more SGSBs than were previously known
(\cite{walter06}). This discovery
means a substantial challenge to
binary star population synthesis models, which try to
reproduce the observed abundances of different types
of binaries.
This source will add to the growing
population of heavily obscured sources.
In addition, this system contributes to tracing the structure
of the scarcely explored Outer Arm of our Galaxy.

\begin{acknowledgements}
  We would like to thank the anonymous referee for the
  valuable suggestions that improved the quality of the paper.
  This work was supported by the Spanish Ministry of Education
  and Science project
  number AYA2010-15431, \emph{De INTEGRAL a IXO: binarias de rayos X y estrellas activas}. Based on observations made with the 
  Italian Telescopio Nazionale Galileo (TNG) operated on the island of
  La Palma by the Fundaci\'on Galileo Galilei of the INAF 
  (Istituto Nazionale di Astrofisica) at the Spanish Observatorio del Roque de
  los Muchachos of the Instituto de Astrof\'{\i}sica de Canarias in Director Discrectionary Time. 
  JJRR acknowledges the support by the
  Spanish Ministerio de Educaci\'on y Ciencia under grant PR2009-0455.
\end{acknowledgements}

\end{document}